\documentclass[aps,prb,twocolumn,superscriptaddress,showpacs,floatfix]{revtex4}
\usepackage{graphicx}

\input{tcilatex}

\begin{document}

\title{Kinetic magnetoelectric effect in a 2D semiconductor strip due to
boundary-confinement induced spin-orbit coupling}
\author{Yongjin Jiang }
\affiliation{Department of physics, Zhejiang Normal University, Jinhua, Zhejiang 321004,
P. R. China}
\author{ Liangbin Hu}
\affiliation{Department of physics and Laboratory of photonic information technology,
South China Normal University, Guangdong 510631, P. R. China }

\begin{abstract}
In a thin strip of a two-dimensional semiconductor electronic system,
spin-orbit coupling may be induced near both edges of the strip due to the
substantial spatial variation of the confining potential in the boundary
regions. In this paper we show that, in the presence of boundary-confinement
induced spin-orbit coupling, a longitudinal charge current circulating
through a 2D semiconductor strip may cause \textit{strong} non-equilibrium
spin accumulation near both edges of the strip. The spins will be polarized
along the normal of the 2DEG plane but in opposite directions at both edges
of the strip. This phenomenon is essentially a kinetic magnetoelectric
effect from the theoretical points of view, but it manifests in a very
similar form as was conceived in a spin Hall effect.
\end{abstract}

\maketitle


\section{Introduction}

There has recently been much interest in a fascinating topic in the research
community, namely \textit{spin Hall effect} ( SHE ). \ SHE is such a
phenomenon that a transverse spin current is generated when a longitudinal
charge current circulates through a sample, and if the sample has a thin
strip geometry, the transverse spin current will cause non-equilibrium spin
accumulation at both edges of the sample\cite%
{Hirsch1999,MurakamiScience2003,SinovaPRL2004}. Such a phenomenon would be
much useful in the context of semiconductor spin-based electronics ( \textit{%
spintronics} ) because it might provide an effective way for generating spin
currents and/or non-equilibrium spin density in a nonmagnetic semiconductors
without use of ferromagnetic metals or ferromagnetic semiconductors, a
principal challenge in semiconductor spintronics\cite%
{Dassarma2004,semibook2002}. From the theoretical points of view, such a
phenomenon can arise from either \textit{intrinsic} spin-orbit (\ SO )
coupling ( i.e., spin-orbit splitting of the band structure )\cite%
{MurakamiScience2003,SinovaPRL2004} or \textit{extrinsic} SO coupling (
i.e., spin-orbit dependent impurity scatterings )\cite{Hirsch1999,DP71} in a
semiconducting material, and correspondingly, the phenomena due to intrinsic
SO\ coupling was termed \textit{intrinsic spin Hall effect} and the
phenomena due to spin-orbit dependent impurity scatterings termed \textit{%
extrinsic spin Hall effect}. From the standpoint of spintronic applications,
intrinsic SHE are more attractive since it is an intrinsic property of a
semiconducting material and does not rely on spin-orbit dependent impurity
scatterings\cite{MurakamiScience2003,SinovaPRL2004}. Recently two
experiments were reported on the observation of SHE. One is on $n$-doped
bulk GaAs by Kato et al.\cite{KatoScience2004} and the other is on
two-dimensional $p$-doped GaAs by Wunderlich et al.\cite{WunderlichPRL2005}.
The phenomenon observed in two-dimensional $p$-doped GaAs by Wunderlich et
al. was believed to be an intrinsic spin Hall effect since the edge spin
accumulation measured in a thin strip of such a system is insensitive to
impurity scatterings in the weak impurity scattering regime\cite%
{WunderlichPRL2005}, an important feature of intrinsic spin Hall effect\cite%
{MurakamiScience2003,SinovaPRL2004}. \ In contrast, the phenomenon observed
in $n$-doped bulk GaAs by Kato et al. was believed to have an extrinsic
origin ( i.e., due to spin-orbit dependent impurity scatterings ) since the
edge spin accumulation measured in a thin strip of such a system is several
order of magnitude smaller compared with the theoretical predictions of
Refs.[2-3]. Although substantial progresses have been achieved on the
detection of spin Hall effect, it should be noted that from the theoretical
points of view there are still intensive debates on whether intrinsic spin
Hall effect does can survive in a spin-orbit coupled system. For example,
several recent theoretical works have argued that, except for the ballistic
transport limit, intrinsic spin\ Hall effect cannot survive in a diffusive
two-dimensional electron gas with Rashba spin-orbit coupling even in the
weak impurity scattering limit\cite%
{InouePRB2004,MishchenkoPRL2004,RashbaPRB2003, Raim05, Dim05, Szhang05,
Kha06}. As to the physical understanding of the recent experimental results%
\cite{KatoScience2004,WunderlichPRL2005}, some significant controversies
also exist which needs further clarifications. For details please refer to
Refs.[16-25].

In this paper, we investigate theoretically another kind of electric-field
driven edge spin-accumulation which might occur in a thin strip of a
two-dimensional electronic ( electron or hole ) system. From the theoretical
points of view, this phenomenon is essentially a \textit{kinetic
magnetoelectric effect}\cite{Edelstein} due to boundary-confinement induced
spin-orbit coupling ( which will be called \textit{edge SO coupling} below
). But very interestingly, it would manifest in a very similar manner as was
conceived in an spin Hall effect. For example, the electric-field driven
edge spin accumulation in a thin 2DEG strip due to this phenomenon will be
polarized perpendicular to the 2DEG plane but along opposite directions at
both edges of the strip; and in the weak impurity scattering regime ( below
a certain disorder strength ) the edge spin accumulation will not decrease
as the disorder strength increases, thus it can survive even in the
diffusive transport regime. These features are very similar to the recently
discovered spin Hall effect, but the mechanisms involved in this phenomenon
are very different from that of the usual spin Hall effect from the
theoretical points of view. The results obtained in the present paper might
provide some new implications to the proper physical understanding of the
recent experimental results\cite{KatoScience2004,WunderlichPRL2005}.

The paper is organized as following: in section II an edge SO coupling model
describing the boundary-confinement induced SO coupling in a thin 2DEG strip
will be introduced and some details of the theoretical formalism used in the
paper will be briefly explained, and in section III some numerical results
will be presented and discussed.

\section{Model and Theoretical Formulation}

The system considered in the present paper consists of a thin 2DEG strip
connected to two ideal leads, as was shown in Fig.1(a). According to the
theory of the relativistic quantum mechanics, if the movement of an electron
is confined by a spatially non-uniform potential, the spin and orbital
degree of freedom of the electron will be coupled together\cite{Merz}. The
longitudinal direction of the strip will be defined as the $x$ direction and
the normal of the 2DEG plane defined as the $z$ direction. For the sake of
simplification, we assume that the confining potential is spatially uniform
along the longitudinal direction of the strip, i.e., the spatial variation
of the confining potential occurs only in the transverse direction of the
strip ( defined as the $y$ direction ). Under this assumption, the SO
coupling due to the transverse spatial variation of the confining potential
will take the following form\cite{Merz, wink},
\begin{equation}
\hat{H}_{SO}=-\frac{\hbar ^{2}}{4m^{2}c^{2}}\hat{\sigma}\times \mathbf{%
k\cdot \nabla }V\mathbf{(}y\mathbf{),}
\end{equation}%
where $\hat{\sigma}=(\hat{\sigma}_{x},\hat{\sigma}_{y},\hat{\sigma}_{z})$
are the usual Pauli matrices, $\mathbf{k}$ is the wave vector of electrons, $%
\mathbf{\nabla }$ denotes the usual gradient operator and $V\mathbf{(}y%
\mathbf{)}$ is the transverse confining potential, which depends only on the
$y$ coordinates. From the effective Hamiltonian (1) one can see that, for an
electron moving along the $+x$ ( or $-x$ ) direction, the SO coupling will
tend to force the electron spin to align along the $+z$ ( or $-z$ )
direction, depending on the sign of $\partial V(y)/\partial y$. In the
absence of external electric field, the numbers of electrons moving along
the $+x$ and $-x$ directions are equal and no net spin density can be
resulted. However, if an external electric field is applied in the
longitudinal direction of the strip, the numbers of electrons moving to $-x$
direction will be larger than the number of electrons moving to the $+x$
direction, and hence a net spin density polarized along the $+z$ or $($ $-z$
$)$ direction might be resulted from the SO coupling. Such an effect was
dubbed the \textit{kinetic magnetoelectric effect} in the literature\cite%
{Edelstein}.
\begin{figure}[tbh]
\includegraphics[width=7cm,height=6cm]{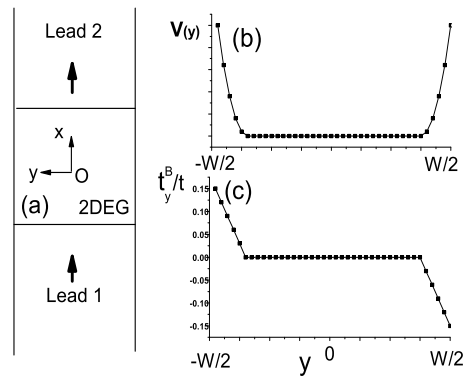}
\caption{(a) A 2D semiconductor strip connected to two ideal leads; (b)
Profile of the transverse spatial variation of the confining potential in
the 2DEG strip; (c) Profile of the transverse spatial variation of the
spin-orbit coupling strength in the 2DEG strip. }
\end{figure}
For a 2DEG strip, if the strip width is much larger than the lattice
constant, one can further assume that the confining potential $V\mathbf{(}y%
\mathbf{)}$ varies substantially only in a narrow boundary region as was
shown in Fig.1(b). ( We assume that the two boundaries of the srip are
located at $y=\pm \frac{W}{2}$ ). Then the SO coupling exists only near both
edges of the strip, and due to the symmetry of the confining potential, the
SO coupling coefficient ( proportional to $\mathbf{\partial }V\mathbf{(}y%
\mathbf{)/}\partial y$ ) has opposite signs at both edges of the strip, as
was shown in Fig.1(c). Thus, the edge SO coupling model described by the
effective Hamiltonian (1) implies that the electric-field driven spin
accumulation will be polarized along opposite directions ( normal to the
2DEG plane ) at both edges of the strip, similar to a spin Hall effect. This
will be confirmed by detailed numerical calculations below. In order to make
a comparison with the widely studied intrinsic spin Hall effect in a Rashba
two-dimensional electron gas, we can also include a Rashba SO coupling in
our edge SO coupling model. Then the total Hamiltonian of the 2DEG strip
will be
\begin{equation}
\hat{H}=\frac{\hbar ^{2}k^{2}}{2m}+\alpha (\hat{\sigma}\times \vec{k})\cdot
\vec{z}-\frac{\hbar ^{2}}{4m^{2}c^{2}}\hat{\sigma}\times \mathbf{k\cdot
\nabla }V\mathbf{(}y)\mathbf{+}V(y)\mathbf{,}
\end{equation}%
where the second term is the Rashba SO coupling ( which arises from the
inversion asymmetry of the trapping well along the normal of the 2DEG plane
) and $\alpha $ the Rashba SO\ coupling constant. ( $\vec{z}$ is a unit
vector along the normal of the 2DEG plane ). A direct use of the Hamiltonian
(2) is not convenient if one wants to take into account the effects of
impurity scatterings properly, which are very important in the diffusive
transport regime. So in our calculations we will transform the Hamiltonian
(2) into a discrete form. The discrete version of the effective Hamiltonian
for the total system ( including both the leads and the 2DEG strip ) will
read:
\begin{widetext}
\begin{eqnarray}
{\cal H}&=&-t\sum_{p=1,2}\sum_{<i,j>\sigma}
(C_{p_{i}\sigma}^{\dag}C_{p_{j}\sigma }+h.c.)+\sum_{R_{i}
}w_{R_{i}}\Psi _{R_{i}}^{\dag }\Psi _{R_{i}}
-t\sum_{<R_{i},R_{j}>}(\Psi _{R_{i}}^{\dag }\Psi _{R_{j}}
 +h.c.)\nonumber\\
&&-t\sum_{p_{n},R_{n}}(C_{p_{n}\sigma }^{\dag }C_{R_{n}\sigma
}+h.c.)-t^{R}\sum_{R_{i}}[i(\Psi _{R_{i}}^{\dag }\sigma ^{x}\Psi
_{R_{i}+y}-\Psi _{R_{i}}^{\dag }\sigma ^{y}\Psi _{R_{i}+x})+h.c.]\nonumber\\
&&-\sum_{R_{i}}t_{R_{i}}^{B}[i\Psi _{R_{i}}^{\dag }\sigma ^{z}\Psi
_{R_{i}+x}+h.c.]\label{eq:one}
\end{eqnarray}
\end{widetext}

Here $t=\frac{\hbar ^{2}}{2m^{\ast }a^{2}}$ is the hopping matrix element
between two nearest-neighbour sites and $a$ the lattice constant in the 2DEG
strip. $\Psi _{R_{i}}=(C_{R_{i},\uparrow },C_{R_{i},\downarrow })$ is the
annihilation operators of electrons at the lattice site $R_{i}$ in the
strip. $C_{p_{j}\sigma }$ is the annihilation operator of electrons with
spin $\sigma $ at the lattice site $p_{j}$ in the lead $p$ ( $p=1,2$ ). $%
(p_{n},R_{n})$ stands for a nearest-neighbor pair of lattice sites across
the interfaces between the 2DEG strip and the leads. $w_{R_{i}}$ is the
on-site energy in the 2DEG strip. In a clean system without disorder, one
usually sets $w_{R_{i}}$ to be zero. $t_{R}=\alpha /2a$ is the Rashba SO
coupling coefficient in the 2DEG strip, which is assumed to be
site-independent. The last term in Eq.~(\ref{eq:one}) stands for the
boundary-confinement induced SO coupling and $t_{R_{i}}^{B}=\frac{\hbar ^{2}%
}{4m^{2}c^{2}}\nabla V$ is the coupling coefficient, which is
site-dependent. The site-dependence of $t_{R_{i}}^{B}$ will be determined by
the actual form of the transverse confining potential in the 2DEG strip. In
our calculations we assume a parabolic confining potential near both edges
of the strip, as was shown in Fig.1(b). In such cases, the site-dependence
of the edge SO coupling coefficient can be expressed as
\begin{equation}
t_{R_{i}}^{B}=\pm t^{B}\max (N_{B}+|y|-N_{y}/2,0),  \label{eq:edgeSO}
\end{equation}%
where $N_{y}=W/a$ is the width of the strip (in units of lattice constant), $%
N^{B}$ is the width of the narrow boundary regions in which the edge SO
coupling exists, and $t^{B}$ is the minimum value of the site-dependent edge
SO coupling coefficient. The sign $\pm $ is different for the two edges, as
shown in Fig.1(c).

Our calculations will be based on the Landauer-Buttiker formula. To this
end, we first consider the transmission and reflection of an electron
incident from a lead. The real space wave function of an incident electron
with spin $\sigma $ will be denoted as $e^{-ik_{m}^{p}x_{p}}\chi _{m\sigma
}^{p}(y_{p})$, where $\chi _{m\sigma }^{p}(y_{p})$ denotes the $m$'th
transverse mode with spin index $\sigma $ in the lead $p$ and $k_{m}^{p}$
the longitudinal wave vector. We adopt the local coordinate scheme for all
leads. In the local coordinate scheme, the longitudinal coordinate $x_{q}$
in the lead $q$ will take the integer numbers $1,2,$...,$\infty $ away from
the 2DEG interface and the transverse coordinate $y_{q}$ take the value of $%
-N_{q}/2,...,N_{q}/2$. The longitudinal wave vector $k_{m}^{p}$ satisfys the
relation $-2t\cos (k_{m}^{p})+\varepsilon _{m}^{p}=E$, where $\varepsilon
_{m}^{p}$ is the eigen-energy of the $m$'th transverse mode in the lead $p$
and $E$ the energy of the incident electron. Including both the incident and
reflected waves, the total wave function in the lead $q$ has the the
following general form:
\begin{eqnarray}
\psi _{\sigma \prime }^{pm\sigma }(x_{q},y_{q}) &=&\delta _{pq}\delta
_{\sigma \sigma \prime }e^{-ik_{m}^{p}x_{p}}\chi _{m\sigma }^{p}(y_{p})
\nonumber \\
&&+\sum_{n\in {q}}\phi _{qn\sigma ^{\prime }}^{pm\sigma
}e^{ik_{n}^{q}x_{q}}\chi _{n\sigma ^{\prime }}^{q}(y_{q})
\label{eq:swavefunction}
\end{eqnarray}%
where $\phi _{qn\sigma ^{\prime }}^{pm\sigma }$ stands for the scattering
amplitude from the ($m\sigma $) mode in the lead $p$ to the ($n\sigma
^{\prime }$) mode in the lead $q$. To obtain the scattering amplitudes $\phi
_{qn\sigma ^{\prime }}^{pm\sigma }$ from the Schr\"{o}dinger equation (
which has now a lattice form and hence there is \textit{a separate equation}
for each lattice site and spin index ), we must solve the wave function $%
\psi _{\sigma ^{\prime }}^{pm\sigma }(R_{i})$ in the 2DEG strip
simultaneously. As dealing with usual scattering problems in quantum
mechanics, we use boundary conditions to determine the scattering amplitudes
$\phi _{qn\sigma ^{\prime }}^{pm\sigma }$. \ In the lattice formalism, the
wave functions in the entire 2DEG strip and in the \emph{first row} of the
leads ( i.e., $x_{q}=1$ ) will be involved in the boundary conditions, which
will be determined from the Schr\"{o}dinger equation. Because Eq.(\ref%
{eq:swavefunction}) is a linear combination of all out-going modes with the
same energy $E$, the Schr\"{o}dinger equation is satisfied automatically in
the lead $q$, except for the lattice sites in the \emph{first row }( i.e., $%
x_{q}=1$ ) of the lead which are connected directly to the 2DEG strip. The
wave function in the \emph{first row of }a lead\emph{\ }( which are
determined by the scattering amplitudes $\phi _{qn\sigma ^{\prime
}}^{pm\sigma }$ ) must be solved simultaneously with the wave function in
the 2DEG strip. To simplify the notations, the wave function in the 2DEG
strip will be defined as a column vector $\psi $ whose dimension is $2N$ ( $%
N $ is the total number of lattice sites in the 2DEG strip ). The scattering
amplitudes $\phi _{qn\sigma ^{\prime }}^{pm\sigma }$ will be arranged as a
column vector $\phi $ whose dimension is $2M$ ( $M$ is the total number of
lattice sites in the first row of the leads ). From the Schr\"{o}dinger
equation for the 2DEG strip and for the first row of a lead, one can show
that the boundary conditions at the interface between a lead and the 2DEG
strip can be expressed as the following forms:
\begin{eqnarray}
\mathbf{A}\psi &=&\mathbf{b}+\mathbf{B}\phi ,  \nonumber \\
\mathbf{C}\phi &=&\mathbf{d}+\mathbf{D}\psi ,  \label{eq:three}
\end{eqnarray}%
where $\mathbf{A}$ and $\mathbf{C}$ are two square matrices with a dimension
of $2N\times 2N$ and $2M\times 2M$, respectively; $\mathbf{B}$ and $\mathbf{D%
}$ are two rectangular matrices describing the hopping interaction between
the leads and the 2DEG strip, whose matrix elements will depend on the
actual form of the geometry of the system. The vectors $\mathbf{b}$\textbf{\
}and\textbf{\ }$\mathbf{d}$ describe the contributions from the incident
waves. The details of these matrices and vectors will be given elsewhere\cite%
{Jiang2}.

After obtaining all scattering amplitudes $\phi _{qn\sigma ^{\prime
}}^{pm\sigma }$ from the boundary conditions, we can calculate the charge
current in each lead through the Landauer-Buttiker formula, $%
I_{p}=(e^{2}/h)\sum_{q}\sum_{\sigma _{1},\sigma _{2}}(T_{p\sigma
_{2}}^{q\sigma _{1}}V_{q}-T_{q\sigma _{1}}^{p\sigma _{2}}V_{p})$, where $%
V_{q}=\mu _{q}/(-e)$ is the voltage applied in the lead $q$ and $\mu _{q}$
is the chemical potential in the lead $q$, $T_{q\sigma ^{\prime }}^{p\sigma
} $ are the transmission probabilities defined by $T_{q\sigma ^{\prime
}}^{p\sigma }=\sum_{m,n}|\phi _{qn\sigma ^{\prime }}^{pm\sigma
}|^{2}v_{qn}/v_{pm}$ and $v_{pm}=2t\sin (k_{m}^{p})$ is the velocity for the
$m$'th mode in the lead $p$. The spin current in each lead can be calculated
similarly, $I_{p}^{\sigma }=-(e/4\pi )\sum_{q}\sum_{\sigma _{2}}[(T_{p\sigma
}^{q\sigma _{2}}-T_{p\bar{\sigma}}^{q\sigma _{2}})V_{q}-(T_{q\sigma
_{2}}^{p\sigma }-T_{q\sigma _{2}}^{p\bar{\sigma}})V_{p}]$. With the wave
function $\psi _{\sigma ^{\prime }}^{pm\sigma }(R_{i})$ in the 2DEG strip at
hand, the non-equilibrium spin density in the 2DEG strip can also be
calculated readily by taking proper ensemble average, and the results can be
expressed as
\begin{equation}
\langle \vec{S}_{\alpha }(R_{i})\rangle =\frac{1}{2\pi }\sum_{pm\sigma }\mu
_{p}/v_{pm}\sum_{\alpha ,\beta }\psi _{\alpha }^{pm\sigma \ast }(R_{i})\vec{%
\sigma}_{\alpha \beta }^{\alpha }\psi _{\beta }^{pm\sigma }(R_{i}),
\end{equation}%
where $\langle \vec{S}_{\alpha }(R_{i})\rangle $ denotes the spin density at
the lattice site $R_{i}$ in the 2DEG strip.

\section{Results and Discussions}

In our calculations we will take the typical values of the electron
effective mass $m=0.04m_{e}$ and the lattice constant $a=3nm$\cite%
{NIttaPRL1997}. The chemical potentials in the leads will be set by fixing
the longitudinal charge current density to the experimental value ( $\approx
100\mu A/1.5\mu m$ ) as reported in Ref.[8]. The Fermi energy of the 2DEG
strip will be set to $E_{f}=-3.8t$ throughout the calculations.
\begin{figure}[tbh]
\includegraphics[width=7cm,height=9cm]{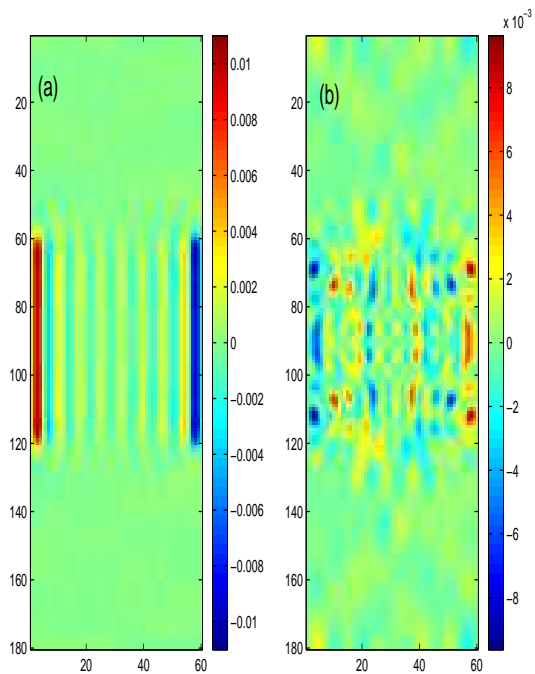}
\caption{The two-dimensional real space configuration of the electric-field
induced non-equilibrium spin density $\langle S_{z}(x,y)\rangle $ (in units
of $\frac{\hbar }{2}$) in the 2DEG strip obtained based on (a) the edge edge
SO coupling model and (b) the Rashba SO coupling model, resepctively. In
Fig.1(a)\ the Rashba SO coupling coefficient $t_{R}$ is set to zero and the
edge SO coupling coefficient $t^{B}$ is set to be $t^{B}=0.03t$. The width
of the boundary regions in which the edge SO coupling exists is set to be $%
N^{B}=5$. In Fig.1(b) the Rashba SO coupling coefficient $t_{R}$ is set to
be $t_{R}=0.15t$ and the edge SO coupling coefficient $t^{B}$ is set to be
zero. In both cases, the 2DEG strip contains $60\times {60}$ lattice sites. }
\end{figure}

In Fig.2(a) we show the typical pattern of the spatial distribution of the
electric-field induced nonequilibrium spin density $<S_{z}>$ in the 2DEG
strip obtained based on the edge SO coupling model. ( Other components of
the spin density is zero in the case of the edge SO coupling model, which
were not shown in the figure ). We have used $180\times 60$ lattice sites in
total ( including both the leads and the 2DEG strip ) in the calculations
and the 2DEG strip contains $60\times 60$ lattice sites. From Fig.2(a) one
can see that, the spatial distribution of the electric-field induced
nonequilibrium spin density in the 2DEG strip obtained based on the edge SO
coupling model manifests in a very similar form as was conceived in a spin
Hall effect\cite{KatoScience2004,WunderlichPRL2005}, i.e., the spins are
polarized perpendicular to the 2DEG plane but along opposite directions on
both edge of the strip and the spin density has two opposite extrema near
both edges. By fixing the longitudinal charge current density to the
experimental value ( $\approx 100\mu A/1.5\mu m$ ) as reported in Ref.[8]
and setting the edge SO coupling coefficient to $t^{B}\simeq 0.03t$, we
found that the spin polarization in the 2DEG strip obtained based on the
edge SO coupling model ( $\approx 1\%$ ) has roughly the same order of
magnitude as the corresponding experimental values reported in Ref.[8]. \ To
make a comparison with the widely studied Rashba SO coupling model\cite%
{NikolicPRL2005}, in Fig.2(b) we have also plotted the spatial distribution
of the electric-field induced nonequilibrium spin density $<S_{z}>$ in the
2DEG strip obtained based on the usual Rashba SO coupling model. From
Fig.2(a) and 2(b) one can see that the typical patterns of the spatial
distribution of the electric-field induced nonequilibrium spin density $%
<S_{z}>$ are very similar in both cases. But it should be pointed out that,
in the case of our edge SO coupling model, only the $z$ component of the
spin density is non zero. In contrast, for the case of the usual Rashba
model, all three components of the spin density are nonzero. ( In Fig.2(b)
we have plotted only the spatial distribution of the $z$ component of the
spin density for comparison ).  Another slight difference that can be seen
from Fig.2(a) and 2(b) is that, the transverse spatial distribution of the
spin density has a very regularly striped pattern in the case of the edge SO
coupling model, but for the case of the usual Rashba model, the spin-density
pattern is not much regularly striped. This slight difference arises from
the fact that, in the case of our edge SO coupling model, the SO coupling
exists only in a narrow boundary region near both edge of the 2DEG strip (
see the illustration shown in Fig.1(b-c) ) and was assumed to be uniform
along the longitudinal direction of the strip. These assumptions lead to a
regularly striped spin-density pattern as shown in Fig.2(a). For the case of
the usual Rashba model, the SO coupling exists in entire the strip ( i.e.,
the SO coupling coefficient is nonzero everywhere in the strip ), thus the
spin-density pattern in the strip is not much regularly striped.
\begin{figure}[tbh]
\includegraphics[width=8cm,height=10cm]{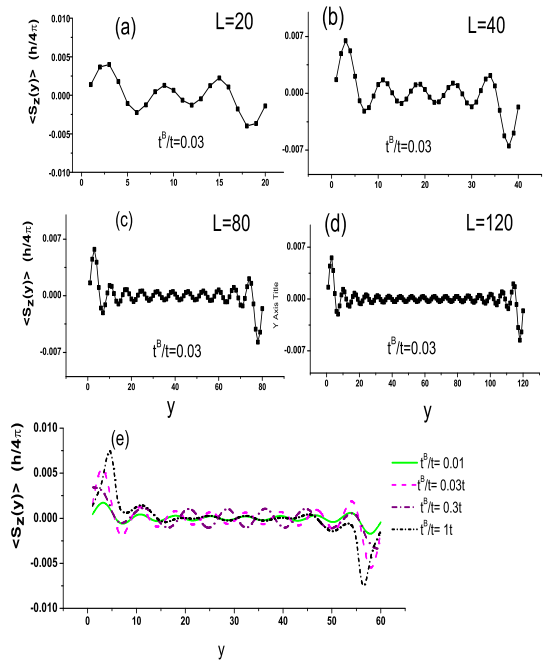}
\caption{The transverse spatial distribution of the longitudinally averaged
spin density $\langle S_{z}(y)\rangle $ in the case of the edge SO coupling
model. The lattice sizes of the 2DEG strip are chosen as: (a) $20\times 20$,
(b) $40\times 40$, (c) $80\times 80$, and (d) $120\times 120$. The edge SO\
couling coefficent is set to be $t^{B}=0.03$. Fig.3(e) shows the dependences
of the transverse spatial distribution of $\langle S_{z}(y)\rangle $ on the
edge SO coupling strength, where the lattice size is fixed to be $60\times
60.$ In all these calculations, the width of the boundary region in which
the edge SO coupling exists is set to be $N^{B}=5$. }
\end{figure}

Next, we study the dependence of the edge spin accumulation on the strip
width and the SO coupling strength in the case of the edge SO coupling
model. Because the spatial distribution of the spin density has a regularly
striped pattern along the longitudinal direction, we can use an averaged
value of $<S_{z}(x,y)>$, defined by $<S_{z}(y)>=\frac{1}{L}%
\int_{0}^{L}S_{z}(x,y)dx$, as a measure of the spin accumulation. In
Fig.3(a-d) we plot the profiles of the transverse spatial distribution of
the spin accumulation obtained in several different cases with different
lattice sizes. From these figures one can see that, $<S_{z}(y)>$ oscillates
inside the strip and has opposite signs on both edges of the strip, and the
magnitude of the spin density will reach a maximum value ( denoted as $%
S_{Max}^{z}$ below ) near both edges of the strip. As the strip width
increases, the transverse spatial distribution of the spin accumulation
become sharper and sharper near both edges of the strip and the oscillations
of the amplitude of the spin density inside the strip tend to be smeared,
i.e., the spin accumulation will be localized near both edges of the strip
if the strip width is much larger than the lattice constant. From these
figures one can also note that, the order of the magnitude of the spin
density near both edges of the strip remains unchanged as the strip width
increases, suggesting that the electric-field induced edge spin accumulation
due to boundary-confinement induced edge SO coupling can survive even in the
diffusive transport regime, similar to the phenomenon reported in Ref.[8]. (
For a 2D semiconductor strip with Rashba spin-orbit coupling, it was
generally believed that the intrinsic spin Hall effect can not survive in
the diffusive transport regime\cite%
{InouePRB2004,MishchenkoPRL2004,RashbaPRB2003, Raim05, Dim05, Szhang05,
Kha06} ). \ The dependence of the transverse spatial distribution of the
spin density on the edge SO coupling strength was shown in Fig.3(e), from
which one can see that the profiles of the transverse spatial distribution
of the spin density do not change substantially as the edge SO coupling
strength varies.

Finally, we discuss the effects of random impurity scatterings on the
electric-field induced nonequilibrium spin density in the case of the edge
SO coupling model. To include properly the effects of random impurity
scatterings, we assume that the on-site energy $\omega _{R_{i}}$ in the 2DEG
strip are randomly but uniformly distributed in an energy region $%
[-W_{D},W_{D}]$, where $W_{D}$ is the amplitude of the on-site energy
fluctuations, which characterizes the disorder strength\cite%
{LShengPRL2005,NikolicPRB2005}. We will calculate the spin density for a
number of random impurity configurations and then do impurity average. In
Fig.4(a-b) we show the relation between the edge spin accumulation and the
disorder strength in both cases of (a) the edge SO coupling model and (b)
the usual Rashba model, respectively. We have done impurity average over
10000 random impurity configurations for each case. From Fig.4(a) one can
see that, for the case of the edge SO coupling model, the edge spin
accumulation does not decrease ( or may even increase ) as the disorder
strength increases in the weak impurity scattering regime ( i.e., below a
certain disorder strength ) and for a fixed longitudinal change current
density ( $\approx 100\mu A/1.5\mu m$ ), similar to the intrinsic spin Hall
effect observed by Wunderlich et al.\cite{WunderlichPRL2005}. In contrast,
for the case of the Rashba model, the edge spin accumulation decreases
monotonously as the disorder strength increases even in the weak impurity
scattering limit, which can be seen clearly from Fig.4(b).
\begin{figure}[tbh]
\includegraphics{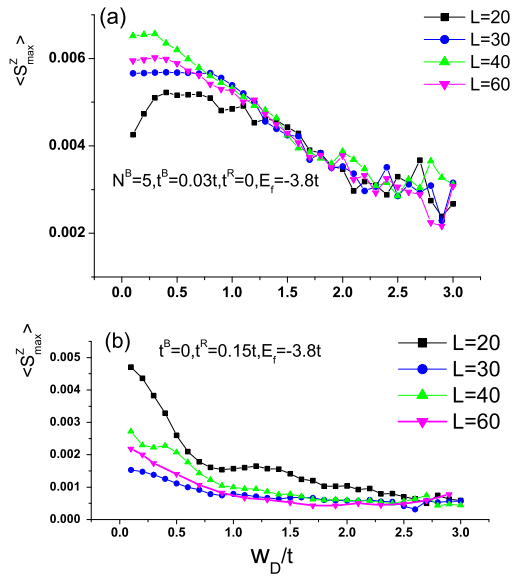}
\caption{The dependences of the edge spin accumulation $\langle
S_{Max}^{z}\rangle $ on the disorder strength $W_{D}$ in both cases of (a)
the edge SO coupling model and (b) the Rashba SO model. In the weak impurity
scattering regime ( below a certain disorder strength ), the edge spin
accumulation \textit{do not} decrease as the disorder strength increases in
the case of the edge SO coupling model. }
\end{figure}
The different behaviors in these two models can be understood qualitatively
as following. For the edge SO coupling model, the SO coupling exists only in
a narrow boundary region near both edges of the strip ( see the illustration
shown in Fig.1(b-c) ). The electric-field induced nonequilibrium spin
density in this model is due to the kinetic magnetoelectric effect but not
due to the flow of a transverse spin Hall current, thus only those
scattering events occurred near both edges of the strip will affect
substantially the spin density. In contrast, for the case of the Rashba
model, the electric-field induced nonequilibrium spin density is due to the
flow of a transverse spin Hall current\cite{NikolicPRL2005}, which will be
damped significantly by all random impurity scattering events occurred in
entire the strip and hence the edge spin accumulation will be decreased
substantially with increasing disorder strength even in the weak impurity
scattering limit. Of course, because localization effects will become
important in the presence of strong impurity scatterings, the electric-field
induced nonequilibrium spin density in the case of the edge SO coupling
model will also be decreased substantially in the presence of strong
impurity scatterings, which can be seen clearly from Fig.4(a).

In summary, we have presented a microscopic model calculation for the
kinetic magnetoelectric effect in a thin strip of a two-dimensional
electronic system due to boundary-confinement induced edge SO coupling. We
have shown that this effect can manifest in a very similar form as was
conceived in a spin Hall effect\cite{KatoScience2004,WunderlichPRL2005}, and
some important features of this effect are similar to the intrinsic spin
Hall effect observed recently in thin strips of two-dimensional \textit{p}%
-doped semiconductors\cite{WunderlichPRL2005}. The results obtained in the
present paper may provide some new implications to the proper physical
understanding of the recent experimental results.

\begin{acknowledgments}
Y. J. Jiang was supported by Natural Science Fundation of Zhejiang province
( Grant No.Y605167 ). L. B. Hu was supported by the National Science
Foundation of China ( Grant No.10474022 ) and the Natural Science Foundation
of Guangdong province ( No.05200534 ).
\end{acknowledgments}

\end{document}